\documentclass[letterpaper,10 pt,conference]{ieeeconf}  

\IEEEoverridecommandlockouts                              
\usepackage{romannum}
\usepackage{cite}
\usepackage{amsmath}
\usepackage{graphicx}
\usepackage{caption}
\usepackage{subcaption}
\usepackage{float}
\usepackage{algorithm}
\usepackage{algorithmic}
\usepackage{xcolor}
\usepackage{amssymb}
\usepackage{textcomp}
\usepackage[utf8]{inputenc}
\usepackage{hyperref}
\usepackage{comment}
\newtheorem{remark}{Remark}
\makeatletter
\newcommand*{\rom}[1]{\expandafter\@slowromancap\romannumeral #1@}
\makeatother

\title{\LARGE \bf
A Deep Reinforcement Learning-based Sliding Mode Control Design for Partially-known Nonlinear Systems*}
\author{Sahand Mosharafian, Shirin Afzali, Yajie Bao, Javad Mohammadpour Velni
\thanks{*This research was financially supported by the United States National Science Foundation under award \#1912757.}
\thanks{S. Mosharafian, S. Afzali, Y. Bao, and J. Mohammadpour Velni are with the School of Electrical \& Computer Engineering, University of Georgia, Athens, GA 30602, USA;
        {\tt\small sahandmosharafian@uga.edu, shirin.afzali@uga.edu, yajie.bao@uga.edu, javadm@uga.edu.}}
}
\begin{document}
\maketitle
\thispagestyle{empty}
\pagestyle{empty}
\begin{abstract}                
Presence of model uncertainties creates challenges for model-based control design, and complexity of the control design is further exacerbated when coping with nonlinear systems. This paper presents a sliding mode control (SMC) design approach for nonlinear systems with partially known dynamics by blending data-driven and model-based approaches. First, an SMC is designed for the available (nominal) model of the nonlinear system. The closed-loop state trajectory of the available model is used to build the desired trajectory for the partially known nonlinear system states. Next, a deep policy gradient method is used to cope with unknown parts of the system dynamics and adjust the sliding mode control output to achieve a desired state trajectory. The performance (and viability) of the proposed design approach is finally examined through numerical examples.
\end{abstract}


\section{Introduction}

\noindent Controller design for nonlinear dynamical systems has been an area of research interest for decades. Various methods for controller design for nonlinear systems have been proposed including feedback linearization \cite{slotine1993robust}, backstepping control \cite{krstic1995nonlinear}, and sliding mode control (SMC) \cite{shtessel2014sliding}. 
Generally, there are plant-model mismatches that arise from parameter uncertainty \cite{9127536}, measurement noise and external disturbances. SMC is a control design technique that offers robustness to these uncertainties in nonlinear systems with stability guarantees \cite{edwards1998sliding, shtessel2014sliding}. However, SMC requires bounds on uncertainties and adds a discontinuity to the system through the $sign$ function, which results in chattering and deteriorates the performance of the SMC. Furthermore, the uncertain knowledge of the system equations would result in a conservative SMC design. Data-driven approaches to control, such as model-free reinforcement learning (RL), require no information about the system and can learn control laws from the data through interactions with system without models \cite{bao21ecc}. However, RL cannot provide stability guarantees and suffers from high sample complexity. In this paper, a reinforcement learning-based SMC design approach is proposed to cope with uncertainties without known bounds by combining the advantages of both RL and SMC. 

RL consists of an agent that interacts with the environment and improves its control actions to maximize the discounted future rewards received from the environment based on the action provided \cite{sutton2018reinforcement}. The distinguishing feature of RL is ``learning by interaction with the environment'' independent of the complexity of the system, thereby enabling RL to be used for complicated control tasks.


There have been recent advancements in the field of artificial intelligence by fusing RL and deep learning techniques. Deep reinforcement learning (DRL) algorithms are resulted from employing deep neural networks to approximate components of reinforcement learning (value function, policy, and model) \cite{ma2019continuous}. Deep Q network (DQN) is a combination of deep neural networks and an RL algorithm called Q-learning which contributed to a significant progress in the fields of games, robotics, and so on \cite{mnih2015human}. However, DQN is only capable of solving discrete problems with low-dimensional action spaces. Therefore, 
In particular, continuous policy gradient methods were proposed to cope with continuous action spaces. Deterministic policy gradient methods \cite{silver2014deterministic} can particularly be useful for controller design applications.

A deterministic policy gradient algorithm based on deep learning and actor-critic is presented in \cite{lillicrap2015continuous}. This method, called deep deterministic policy gradient (DDPG), can handle continuous and high-dimensional action spaces and is used in this paper to design a sliding mode controller. DDPG is an actor-critic, model-free, off-policy algorithm, in which critic learns the Q-function using off-policy data, and actor learns the policy using the sampled policy gradient \cite{lillicrap2015continuous}.

A number of previous studies employed RL for designing SMC. 
Authors in \cite{fan2015adaptive} estimated the uncertainties and disturbance terms respectively by an NN approximator and a disturbance observer for the SMC integrated with RL.  
Moreover, \cite{zhang2018optimal} proposed optimal guaranteed cost SMC integrated with the approximate dynamic programming (ADP) algorithm based on a single critic neural network (NN) for constrained-input nonlinear systems with disturbances. Different from the existing works, in our work, we assume the system is partially known and the goal is to achieve a desired performance for the original system  using the knowledge of a simplified model of the system. In particular, we present an RL-based SMC design approach which preserves the structure of the SMC law by combining the SMC designed for the nominal model and the RL for coping with uncertainties. Instead of using fixed bounds for SMC, the proposed approach can cope with time-varying (and even state- and input-dependent) uncertainties by virtue of the model-free off-policy policy gradient RL algorithm.      

The novelty of our work reported in this paper lies in \textit{\textbf{fusing model-based and data-driven approaches}} for the design of an SMC for a class of nonlinear systems. The model-based part of the controller is obtained through available knowledge about the nonlinear system dynamics. The data-driven part of the controller is then calculated using DDPG algorithm to cope with the discrepancy between the original system and the available model of the system. Moreover, no information about the unknown parts of the system dynamics is needed, and the desired performance is reached. 
Furthermore, the control input, as well as the system states are penalized when defining the reward function for the RL agent to limit chattering. It is noted that \textit{since the plant-model mismatch is used by DDPG to update the SMC output, the proposed design approach interacts with the actual system online and hence leads to less conservative results compared to the traditional robust SMC design methods in the literature}.

The remainder of this paper is organized as follows. Preliminaries and problem statement are provided in Section \rom{2}. Section \rom{3} describes the SMC design process. Section \rom{4} discusses the DDPG algorithm for SMC design purposes. Simulation results are presented in Section \rom{5} to validate the performance of the proposed design approach, and concluding remarks are provided in Section \rom{6}.

\section{Problem Statement and Preliminaries}
\noindent This section first presents the model of the system under study and then provides a brief description of the policy gradient method in reinforcement learning.

\subsection{System Model}

Consider a class of nonlinear systems with $n$ measurable states described in the normal form as \vspace{+0pt}
\begin{equation}
    \begin{gathered} \label{original_sys}
    \dot{x}_i(t) = x_{i+1}(t) \, ; \,\,~~~ i=1,2,\hdots, n-1,\\
    \dot{x}_n(t) = f(t,\mathbf{x}(t)) + \Delta f(t,\mathbf{x}(t)) \\
    + \left[g(t,\mathbf{x}(t)) + \Delta g(t,\mathbf{x}(t)) \right]\,u(t),
    \end{gathered}
\end{equation}
where $\mathbf{x}(t)\in\mathbf{R}^{n}$ is the vector of all system states, $u(t)\in\mathbf{R}$ is the control input, $f(t,\mathbf{x}(t)) \in\mathbf{R}$, and $g(t,\mathbf{x}(t))\in\mathbf{R}$. Assume that $\Delta f(t,\mathbf{x}(t))$ and $\Delta g(t,\mathbf{x}(t))$ are unknown. A simplified model of the original system in \eqref{original_sys} can be represented as follows \vspace{-2pt}
\begin{equation}
    \begin{gathered} \label{simplified_sys}
    \dot{\hat{x}}_i(t) = \hat{x}_{i+1}(t) \, ; \,\,~~~~ \,\,\,i=1,2,\hdots, n-1,\\
    \dot{\hat{x}}_n(t)=f(t,\mathbf{\hat{x}}(t)) + g(t,\mathbf{\hat{x}}(t))\,\hat{u}(t),
    \end{gathered}
\end{equation}
where $\mathbf{\hat{x}}(t)$ is the vector of all simplified system states. The goal is to design an RL-based sliding mode controller (SMC) with partial knowledge of the system dynamics (here, the partial knowledge is the simplified system model). It is noted that the simplified system can be even considered to be a linear approximation of the original system. 

\begin{remark}
The original system model can be 
described in the strict feedback form
\begin{equation}
    \begin{gathered} \label{original_sys_strict_feedback}
    \dot{x}_1 = f_1(t,x_{1}(t)) + g_1(t,x_{1}(t))\,x_2,\\
    \dot{x}_2 = f_2(t,x_{1}(t),x_{2}(t)) + g_2(t,x_{1}(t),x_{2}(t))\,x_3,\\
    \vdots\\
    \dot{x}_{n-1} = f_{n-1}(t,x_{1}(t),x_{2}(t),\hdots,x_{n-1}(t))\\ \hspace{55 pt}+g_{n-1}(t,x_{1}(t),x_{2}(t),\hdots,x_{n-1}(t))\,x_n,\\
    \dot{x}_n=f_n(t,\mathbf{x}(t)) + \Delta f(t,\mathbf{x}(t)) \\
    \hspace{55 pt} + \left[g_n(t,\mathbf{x}(t)) + \Delta g(t,\mathbf{x}(t)) \right]\,u(t),
    \end{gathered}
\end{equation}
where the uncertainties are assumed to only exist in the expression of $\dot{x}_{n}$.
The simplified model of the system described by \eqref{original_sys_strict_feedback} will also be in the strict feedback form. 
However, the strict feedback form can be transformed into the normal form using a state transformation \vspace{-5pt}
\begin{equation}\label{state-trans}
    z_{1}=x_{1};z_{2} = \dot{z}_{1};\cdots;z_{n} = \dot{z}_{n-1}
\end{equation}
where $z_{n}$ requires no information about $\dot{x}_{n}$, and the simplified model can be transformed similarly. Therefore, the RL-based SMC design approach proposed in this paper can be extended to treat systems in the form of  (\ref{original_sys_strict_feedback}).
\end{remark}

\subsection{Policy Gradient in Reinforcement Learning}

A reinforcement learning (RL) agent aims at learning a policy that maximizes the discounted future rewards (expected return). The return at time step $t$ is the total discounted reward from $t$ as $G_t=\sum_{k=t}^{N}\gamma^{k-t}\,r_k$, where $r_k=r(s_k,a_k)$ is the reward received by taking action $a_k$ in state $s_k$, and $0<\gamma\leq 1$ is the discount rate. For non-episodic tasks, $N$ is $\infty$. The value function evaluates the expected return beginning from state $s$ under policy $\pi$, and represented as $V^{\pi}(s)=\mathbf{E}_{\pi}[G_t|S_t=s]$. The expected return beginning from state $s$ and taking action $a$ is defined as Q-value $(Q^{\pi}(s,a)=\mathbf{E}_{\pi}[G_t|S_t=s,A_t=a])$ following policy $\pi$. The RL agent aims at maximizing its expected return beginning from the initial state; thus, the agent's goal is to maximize $J(\pi)=V^\pi(s_0)=E_{\pi}[G_0]$.

In policy gradient algorithms, which are suitable for RL problems with continuous action space \cite{silver2014deterministic}, the policy is parametrized by additional sets of parameters $\mathbf{\theta}$, which can be the weights of a neural network ($\pi(s,\mathbf{\theta})=\pi_\mathbf{\theta}(s)$). In this case, the objective function for RL agent turns into $J(\pi_{\theta})=E_{\pi_{\theta}}[G_0]$. In policy gradient algorithms, the goal is to update policy parameters $\theta$ to maximize $J$; hence, the parameters $\theta$ are updated in the direction of $\nabla_{\theta}J$. In \cite{sutton2018reinforcement}, it is shown that for stochastic policies
 \begin{equation}
     \begin{gathered}
     \nabla_{\theta}J(\pi_{\theta}) = \int_S \rho(s) \int_A \nabla_{\theta}\,\pi_{\theta}(a|s)\,Q^{\pi}(s,a)\,\mathrm{d}a\,\mathrm{d}s\\
     = \mathbf{E}_{s\sim\rho^{\pi},\,a\sim\pi_{\theta}}[\nabla_{\theta}\,\mathrm{log}\,\pi_{\theta}(a|s)\,Q^{\pi}(s,a)],
     \end{gathered}
 \end{equation}
where $\rho(s)$ is the state distribution following policy $\pi_{\theta}$.

Actor-critic algorithms, which use policy gradient theorem, consist of an actor which adjusts policy parameters $\theta$, and a critic which estimates $Q^{\pi}(s,a)$ by $Q^{\phi}(s,a)$ with parameters $\phi$ \cite{grondman2012survey}. The critic tries to adjust parameters $\phi$ in order to minimize the following mean squared error (MSE)
\begin{equation}
    \begin{gathered}
    L(\phi) = \mathbf{E}_{s\sim\rho^{\pi},\,a\sim\pi_{\theta}}\left[(\,Q^{\phi}(s,a)-Q^{\pi}(s,a)\,)^{2}\right].
    \end{gathered}
\end{equation}
For designing controllers using policy gradient in this paper, continuous deterministic policy is used, and the gradient of policy should be adapted to improve the deterministic policy. According to \cite{silver2014deterministic}, in policy improvement methods, a common approach to update policy is to find a greedy policy such that
\begin{equation*}
    \mu^{k+1}(s)=\mathrm{arg}\max_{a}Q^{\mu^{k}}(s,a).
\end{equation*}
The notation $\mu(s)$ is used to show the deterministic policy. Since the greedy policy improvement is computationally expensive for continuous action spaces, the alternative method for improving the parametrized policy is to move in the direction of $\nabla_{\theta}Q^{\mu^{k}}(s,\mu_{\theta}(s))$. Hence, the updating formula for improving policy is represented as \cite{silver2014deterministic}
\begin{equation} \label{update_theta}
    \theta^{k+1}=\theta^{k+1}+\alpha_a\,\mathbf{E}_{s\sim\rho^{\mu^{k}}}\left[\nabla_{\theta}Q^{\mu^{k}}(s,\mu_{\theta}(s))\right],
\end{equation}
where $\alpha_a$ is the learning rate. It is shown in  \cite{silver2014deterministic} that \begin{equation}
    \nabla_{\theta}J(\mu_{\theta})=\mathbf{E}_{s\sim\mu^{k}}\left[\nabla_{\theta}Q^{\mu^{k}}(s,\mu_{\theta}(s))\right],
\end{equation}
which implies that the update formula \eqref{update_theta} moves policy parameters $\theta$ in the direction that maximizes $J(\mu_{\theta}(s))$. The update formula is used later in the paper to find a suitable control action for the original system \eqref{original_sys}.
\section{Design of an SMC for the Original System}

\noindent Since $\Delta f(t,\mathbf{x}(t))$ and $\Delta g(t,\mathbf{x}(t))$ as well as their bounds are unknown, designing a controller for the original system is not straightforward. 
First, an SMC for the simplified system is designed to use the existing knowledge. Then, RL is used to cope with the uncertainties in the original system while preserving the structure of the sliding mode controller.

To design SMC for the simplified system, by defining a stable sliding surface as \vspace{-5pt}
\begin{equation}\label{simplified_sliding_surface}
    \hat{\sigma}(\hat{\mathbf{x}}) = \sum_{i=1}^{n}a_i\,\hat{x}_i\,;\,\,\,\,\,\,a_i>0,\,\,\, \,i=1,2,\hdots,n,
\end{equation}
the controller is \vspace{-5pt}
\begin{equation}
\begin{gathered} \label{u_hat}
    \hat{u}(t,\hat{\mathbf{x}})= \hat{u}_c(t,\hat{\mathbf{x}}) + \hat{u}_{eq}(t,\hat{\mathbf{x}}),\\
\end{gathered}
\end{equation}
where \vspace{-5pt}
\begin{equation}
\begin{gathered} \label{u_c,eq}
    \hat{u}_{eq}(t,\hat{\mathbf{x}}) =\frac{-1}{a_n\,g(t,\mathbf{\hat{x}})} \left[
    \sum_{i=1}^{n-1}a_i\,\hat{x}_{i+1} +a_n\,f(t,\mathbf{\hat{x}}) \right],\\
    \hat{u}_{c}(t,\hat{\mathbf{x}})= \frac{-\hat{\mu}}{a_n\,g(t,\mathbf{\hat{x}})}\, sign(\hat{\sigma}).
\end{gathered}
\end{equation}
The error is defined as \vspace{-5pt}
\begin{equation}
    e_i(t) = x_i(t)-\hat{x}_i(t);\,\,\, \,\,\, i=1,2,\hdots,n.
\end{equation}
Therefore, the error system is a nonlinear system in the normal form as
\begin{equation}
    \begin{gathered} \label{error_sys}
    \dot{e}_i(t) = e_{i+1}(t) \, ; \,\, \,\,\,i=1,2,\hdots, n-1,\\
    \dot{e}_n(t)=f(t,\mathbf{x}) + \Delta f(t,\mathbf{x}) - f(t,\mathbf{\hat{x}})\\
    + \left[g(t,\mathbf{x}) + \Delta g(t,\mathbf{x}) \right]u(t,{\mathbf{x}})- g(t,\hat{\mathbf{x}})\hat{u}(t,\hat{\mathbf{x}}).
    \end{gathered}
\end{equation}

Now, a new stable sliding surface is defined for the error system as $\sigma = \sum_{i=1}^{n}a_i\,e_i$. 
The first-order derivative of $\sigma$ is \vspace{-5pt}
\begin{equation}
\begin{gathered} \label{sigma_dot}
    \dot{\sigma} = \sum_{i=1}^{n-1}a_i\,e_{i+1} + a_n\,\Big[f(t,\mathbf{x}) + \Delta f(t,\mathbf{x}) - f(t,\mathbf{\hat{x}}) \\
    + \left[g(t,\mathbf{x}) + \Delta g(t,\mathbf{x}) \right]u(t,{\mathbf{x}})- g(t,\hat{\mathbf{x}})\hat{u}(t,\hat{\mathbf{x}})\Big].
    \end{gathered}
\end{equation}

With the control law \eqref{u_hat} for the simplified model, we consider the controller for the original system in the form of \vspace{-5pt}
\begin{equation} \label{u(t)}
    u(t,{\mathbf{x}})=\hat{u}(t,\mathbf{x})+u_1(t),
\end{equation}
where $u_{1}$ is used to compensate for the plant-model mismatch and will be learned by RL. By substituting $\hat{u}$ and $u$ in \eqref{sigma_dot} with \eqref{u_hat} and \eqref{u(t)}, $\dot{\sigma}$ turns into \vspace{-2pt}
\begin{equation}
    \begin{gathered}\label{original_sys_sliding_surface_diff}
        \dot{\sigma}=        {\hat{\mu}\left[\, sign(\hat{\sigma}(\hat{\mathbf{x}}))-sign(\hat{\sigma}({\mathbf{x}}))\right]}\hspace{95pt}\\\hspace{10pt}
        -\frac{\Delta g(t,\mathbf{x})}{g(t,\mathbf{x})}\left[a_n\,f(t,\mathbf{x})+\sum_{i=1}^{n-1}a_i\,x_{i+1}+\hat{\mu}\,sign(\hat{\sigma}(\mathbf{x}))\right]\\
        +a_n\,\Delta f(t,\mathbf{x})
        +a_n\,\left[g(t,\mathbf{x})+\Delta g (t,\mathbf{x})\right]u_1(t).\hspace{25pt}
    \end{gathered}
\end{equation}

To design an SMC for the error system, $u_1(t)$ is chosen as \vspace{-5pt}
\begin{equation} \label{u_1}
    u_1(t)=-r(t) - \mu(t)\,sign(\sigma),
\end{equation}
where $r(t)$ and $\mu(t)$ need to be designed. 
Ideally, consider a Lyapunov function candidate $V = \frac{1}{2} \sigma^{\top}\sigma$, \vspace{-2pt}
\begin{equation}
\label{eq:opt-u}
\begin{split}
    r(t) &= \frac{1}{a_n\,\left[g(t,\mathbf{x})+\Delta g (t,\mathbf{x})\right]}\Bigg\{\frac{-\Delta g(t,\mathbf{x})}{g(t,\mathbf{x})}\bigg[a_n\,f(t,\mathbf{x}) \\
    &+\sum_{i=1}^{n-1}a_i\,x_{i+1}+\hat{\mu}\,sign(\hat{\sigma}(\mathbf{x}))\bigg]+a_n\,\Delta f(t,\mathbf{x}) \\
    &+\hat{\mu}\left[\, sign(\hat{\sigma}(\hat{\mathbf{x}}))-sign(\hat{\sigma}({\mathbf{x}}))\right]\Bigg\}, \\
\mu(t) &= \frac{1}{a_n\,\left[g(t,\mathbf{x})+\Delta g (t,\mathbf{x})\right]}sign(\sigma)
\end{split}
\end{equation} 
such that $\dot{V} = \sigma^{\top}\dot{\sigma}=-\sigma^{\top}sign(\sigma) \leq 0$, which guarantees $\sigma\xrightarrow[]{}0$ in finite time. Also, \eqref{eq:opt-u} shows the deficiency of the nominal controller \eqref{u_hat}.   

If the original system does not have any unknown parts ($\Delta f\rightarrow0$ and $\Delta g \rightarrow 0$), and $u_1(t)= 0$, then $\mathbf{x}\rightarrow\mathbf{\hat{x}}$ and nothing is left to be designed. However, the original system model is not completely available. The desired case is when the simplified system is close to the original system, but it might not be always the case. When $\Delta f$ and $\Delta g$ are large, $u_1(t)$ is significant and $\hat{u}(t)$ alone will not achieve the control objective with stability guarantees. To learn $u_1(t)$ in the form of \eqref{u_1} without the knowledge of $\Delta f$ and $\Delta g$, DDPG is used, which is elaborated in the next section.

\begin{remark}\label{remark:2}
For a tracking controller design problem using sliding mode for the original system represented in \eqref{original_sys}, where the desired output for the first state is $y(t)=x_1^{ref}(t)$, first a new system based on the simplified system (equation \eqref{simplified_sys}) and the desired output is defined by assuming
$\hat{e}_1(t) = \hat{x}_1(t)-y(t)$. Hence, this system dynamics turn into
\begin{equation}
    \begin{gathered}\label{simplified_tracking_sys}
    \dot{\hat{e}}_1 = \hat{e_2} = \hat{x}_2-\dot{y}\\
    \dot{\hat{e}}_2 = \hat{e_3}=\hat{x}_3-\Ddot{y}\\
    \vdots\\
    \dot{\hat{e}}_{n-1} = \hat{e_n}=\hat{x}_n-y^{(n-1)}\\
    \dot{\hat{e}}_{n}= -y^{(n)}+f(t,\mathbf{x}(t)) + \Delta f(t,\mathbf{x}(t)) \\
    + \left[g(t,\mathbf{x}(t)) + \Delta g(t,\mathbf{x}(t)) \right]\,u(t).
    \end{gathered}
\end{equation}
Then, the procedure described above for designing SMC for the original system can be employed by replacing $\hat{x}_i$ with $\hat{e}_i$ in equations \eqref{simplified_sliding_surface}-\eqref{original_sys_sliding_surface_diff}.
\end{remark}
\color{black}

\section{Deep Reinforcement Learning Controller Design for the Error System}

\noindent Deep deterministic policy gradient (DDPG) algorithm was introduced in \cite{lillicrap2015continuous}. In this method, actor learns a deterministic policy while critic learns the Q-value function. Since the Q-value update may cause divergence \cite{lillicrap2015continuous}, a copy of the actor network and a copy of the critic network are considered as target networks. DDPG uses soft target updates instead of directly copying weights from the original network. Hence, target network weights are updated slowly based on the learned network. Although soft target update may slow down the learning process, its stability improvement outweighs the low learning speed. A major challenge in deterministic policy gradient methods is exploration; adding noise to the deterministic policy can improve the exploration and avoid sub-optimal solutions \cite{lillicrap2015continuous}. The added noise can be an Ornstein-Uhlenbeck process \cite{uhlenbeck1930theory}.
Based on \cite{lillicrap2015continuous}, in DDPG algorithm, the critic updates Q network weights to minimize the following loss \vspace{-5pt}
\begin{multline}
    L = \mathbf{E}_{s\sim\rho^{\mu}}\big\{[\,\,Q^{\phi}(s,\mu_{\theta}(s))-(\,r(s,\mu_{\theta}(s))+\\  \gamma Q^{\phi_{t}}(s',\mu_{\theta_{t}}(s'))\,)\,\,]^{2}\big\},
\end{multline}
where $\phi$ represents Q network parameters, while $\phi_{t}$ and $\theta_{t}$ are target Q network and target actor network parameters, respectively. Sample-based loss can be simply calculated by \vspace{-5pt}
\begin{multline}
    L_{B_1}=\frac{1}{|B_1|}
    \sum_{(s,\mu_{\theta}(s),r,s') \in B_1}[\,\,Q^{\phi}(s,\mu_{\theta}(s))
    -(\,r(s,\mu_{\theta}(s))\\+\gamma \, Q^{\phi_{t}}(s',\mu_{\theta_{t}}(s'))\,)\,\,]^{2},
\end{multline}
where $B_1$ is a mini-batch of the sampled data and $|B_1|$ is the number of samples in the mini-batch.

\begin{figure}[!htp]
\begin{center}
\includegraphics[width=1\linewidth]{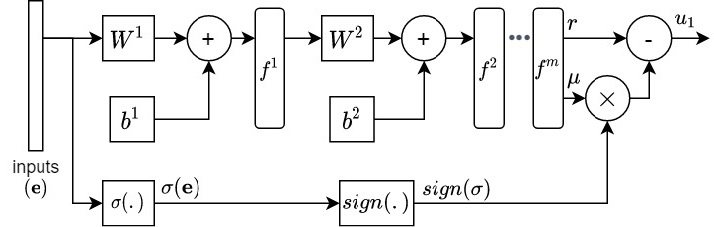}    
\caption{Actor network diagram: the custom output layer is designed to create $u_1$. Trainable weights of the network (which build policy parameters $\theta$) are $W^i$ and $b^i$, $f^i$ is the activation function.} 
\label{fig:actor network}
\vspace{5pt}
\includegraphics[width=1\linewidth]{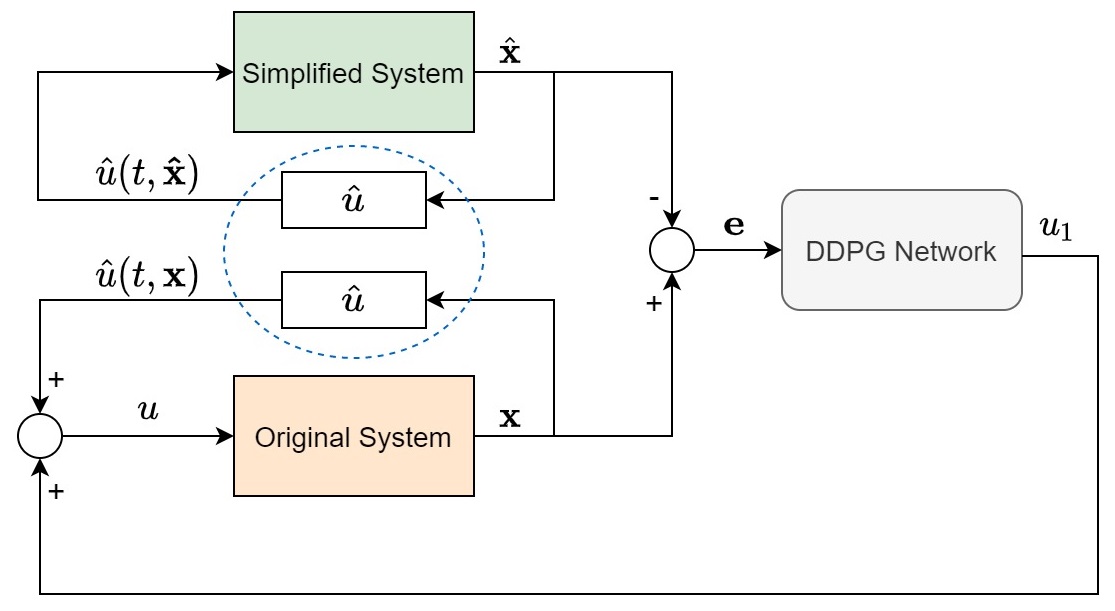}    
\caption{Closed-loop configuration: DDPG network uses error signal to find $u_1$ to maximize the reward (minimize the weighted square sum of the error system states and the input).}
\label{fig:diagram}
\end{center}
\vspace{-5pt}
\end{figure}

For designing SMC, the structure considered for the actor network is shown in Fig. \ref{fig:actor network}. The output layer is customized to achieve the desired form of control signal given in \eqref{u_1}. Based on Fig. \ref{fig:actor network}, the activation functions of the last layer (before the custom layer) generate two outputs; the one that generates $r$ is linear, while for generating $\mu$, tangent hyperbolic (tanh) activation function is used. Rectified linear activation function is used for the rest of the layers. It is assumed that no information is available about the sign of $g+\Delta g$, and $\mu$ might be positive or negative; since the output of tanh function is between -1 and 1, $\mu$ is bounded between -1 and 1. Bounds on $\mu$ result in limited chattering of the control signal (in case larger bounds on $\mu$ are needed, $\mu$ can be multiplied by a fixed number). If the sign of $g+\Delta g$ does not change in a vicinity of the origin, then tanh activation function can be replaced by sigmoid function.

The structure of the closed-loop system controlled by DDPG controller is shown in Fig. \ref{fig:diagram}. The DDPG network uses $\mathbf{e}$ as input and generates $u_1$ as its output. For DDPG to learn the optimal control signal, the performance index needs to be defined. By sampling every $t_s$ seconds from the original and the simplified system states, the objective would be for the DDPG algorithm to maximize the following cost function \vspace{-5pt}
\begin{equation} \label{J}
\begin{gathered}
    \max_{\theta}\sum_{k=0}^{N-1}\sum_{i=1}^{n}-q_i\,[e_i(k)]^2-q_u\,[u_1(k)]^{2},
\end{gathered}
\end{equation}
where $u_1=\mu_{\theta}(s)$, $q_i>0$ is a weight indicating the importance of the error system state $e_i$ in the optimization problem, $q_u \geq0$ penalizes $u_1$ (i.e., large control efforts and hence high amplitude chattering), and $N$ denotes the episode length. Based on the defined objective function, the reward at step $k$ of each episode is simply considered as $r(\mathbf{e},\mu_{\theta}(\mathbf{e}))=\sum_{i=1}^{n}-a_i^{2}\,[e_i(k)]^2-b\,[u_1(k)]^{2}$. It is noted that when $t_s \rightarrow 0$, the summation over $k$ in \eqref{J} turns into an integral, and the optimal solution results from solving the HJB equations \cite{bertsekas2000dynamic}. 

The SMC design procedure for the nonlinear error system \eqref{error_sys} is summarized as follows.
\vspace{3mm}
\begin{algorithmic}[1]
\STATE $\mathbf{procedure}$: SMC design for partially-known nonlinear systems
\STATE Input: initial policy network parameters $\mathbf{\theta}$, Q-learning network parameters $\mathbf{\phi}$, empty replay buffer $B$, episode length $N$, number of total episodes for training $N_{ep}$, bound on the reward at each step $\underline{G}$, initial system states $S_0$, learning rates $\alpha_c$, $\alpha_a$, and $\tau$.
\WHILE{$counter<N_{ep}$}
\STATE reset the system ($s\leftarrow s_0$)
\STATE $counter\leftarrow 0$
\WHILE{the episode is not terminated}
\STATE select action $u_1$ based on current state $s$
\STATE apply $u=\hat{u}(\mathbf{x})+u_1$ to the original system
\STATE observe next state $S'$ and reward $R$, and store $(s,s',u_1,r)$ in replay buffer $B$
\STATE sample a mini-batch $B_1$ from $B$
\STATE update Q network parameters using\\ $\phi\leftarrow \phi -\alpha_c \nabla_{\phi}L_{B_1}$
\STATE update policy network using\\
$\theta\leftarrow\theta+\alpha_a\,\nabla_{\theta}\frac{1}{|B_1|}\sum_{s \in B_1}Q_{\phi}(s,\pi_{\theta}(s))$
\STATE update target networks using\\
$\theta_t\leftarrow\tau\,\theta+ (1-\tau)\,\theta_t$\\
$\phi\leftarrow\tau\,\phi+ (1-\tau)\,\phi_t$
\STATE $counter\leftarrow counter+1$
\IF {$counter \geq N$ or $return<\underline{G}$}
\STATE the episode is terminated
\ENDIF
\ENDWHILE
\ENDWHILE
\STATE $\mathbf{end\,procedure}$
\end{algorithmic}

\vspace{3mm}

\begin{remark}
It is noted that the data-driven component of the proposed SMC action does not need information about bounds on the uncertain parts of the system model. Instead, it learns to utilize the discrepancy between the simplified model and the original system through interaction with the closed-loop system. Therefore, the performance of the proposed SMC is less conservative compared to traditional robust SMC design approaches in the literature that only use bounds on the model uncertainties.
\end{remark}

\section{Simulation Results and Discussion}

\noindent To evaluate the performance of the proposed RL-based sliding mode controller design approach, a nonlinear spring-mass-damper system is used. 

\vspace{2mm}

\subsubsection{Case description}
\begin{figure}
\begin{center}
\includegraphics[width=\linewidth]{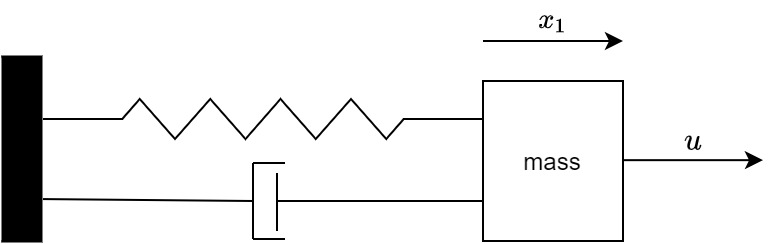}
\caption{A mass-spring-damper system.} 
\label{fig:example}
\end{center}
\vspace{+1mm}
\end{figure}

The state-space representation of the nonlinear mass-spring-damper shown in Fig. \ref{fig:example} (the original system) is as follows \vspace{-5pt}
\begin{equation}
    \begin{gathered}\label{example_original}
    \dot{x}_1=x_2,\\
    \dot{x}_2 =\dfrac{1}{m}[ -c\,{x}_2|{x}_2|- k\,x_1 - b\,x_1^{3} + u],
    \end{gathered}
\end{equation}
where $m$ is the mass, $c$ is the damping coefficient for the nonlinear damper, and $k$ and $b$ represent the nonlinear spring parameters. The available model of the system is, however, a linear system (i.e., the simplified system) derived based on the available knowledge of the physical system with the following differential equation:
\begin{equation}
    \begin{gathered}\label{example_simplified}
    \dot{\hat{x}}_1 = \hat{x}_2,\\
    \dot{\hat{x}}_2 =\dfrac{1}{\hat{m}}[ -\hat{c}\,\hat{x}_2- \hat{k}\,\hat{x}_1 + \hat{u}].
    \end{gathered}
\end{equation}

The constant values in equations \eqref{example_original} and \eqref{example_simplified} are given in Table \ref{table:2}. Our goal is to solve a tracking control problem using the proposed RL-based sliding mode control design method. According to Remark \ref{remark:2} and the design process explained in the previous sections, the control law for tracking $x_1^*(t)=y(t)=\sin(t)-1$ of the simplified system turns into \vspace{-5pt}
\begin{equation}
    \begin{gathered}
    \hat{u}(t,\hat{x}_1,\hat{x}_2)=\hat{m}[\hat{x}_2+\cos(t)-\sin(t)+\dfrac{\hat{k}}{\hat{m}}\hat{x}_1+\dfrac{\hat{c}}{\hat{m}}\hat{x}_2]\\
    \hspace{-65pt}-\,\hat{m}\,[sign(\hat{\sigma})],
    \end{gathered}
\end{equation}
when the following sliding surface is used: \vspace{-5pt}
\begin{equation*}
    \hat{\sigma}=\hat{e}_1 + \hat{e}_2\\
    =\hat{x}_1+\hat{x_2}+1-\sin(t)-\cos(t).
\end{equation*}
Then, the controller for the original system turns into \vspace{-5pt}
\begin{equation*}
    u=\hat{u}(t,x_1,x_2)-r(t)-\mu(t)\,sign(\sigma),
\end{equation*}
where $\sigma=e_1+e_2$, and DDPG will be employed to learn $r(t)$ and $u(t)$.

\vspace{+1mm}
\begin{table}
\caption{Mass-spring-damper system parameters}
\centering
\begin{tabular}{|p{1.25cm}|p{1.8cm}||p{1.25cm}|p{1.8cm}|} 
 \hline
 parameter & value & parameter & value\\ [0.5ex] 
 \hline\hline
 $m$ & $0.8\,kg$
 & $\hat{m}$ & $1\,kg$ \\ 
 $c$ & $2.2\,Ns/m$ & $\hat{c}$ & $2\,Ns/m$\\
 $k$ &  $5.5\,N/m$ & $\hat{k}$ & $5\,N/m$ \\
 $b$ & $1.5\,N/m^3$ & &\\
 [1ex] 
 \hline
\end{tabular}
\label{table:2}
\end{table}
\vspace{1mm}

\vspace{2mm}

\subsubsection{Implementation of DDPG}
For implementing DDPG, Keras package \cite{chollet2015keras} is used. For implementing the proposed control law, two networks with the same structure are used as the actor and its target. These networks consist of 6 layers. The output layer structure is customized to build the desired form of control signal $u_1$ as in \eqref{u_1} (shown in Fig. \ref{fig:actor network}). Each of the first three layers includes 512 units with rectified linear activation function, while the fourth layer includes 64 units with linear function. The fifth layer includes 2 units and the last layer (output layer) is customized as shown in Fig. \ref{fig:actor network}. The inputs to the networks are the error system states. The critic network and its target network are identical and divided into two parts; the first part with error system states as inputs consists of three 512-unit hidden layers. The second part also includes three 512-unit hidden layers but the input of this part is the output of the actor network. Then, the last layer of these two parts are concatenated and the concatenated output is connected to two 512-unit hidden layers. Finally, the output layer builds a single output. Ornstein–Uhlenbeck process with standard deviation of $\sigma=0.1$ is added to the output of the actor network during the learning for exploration.

\subsubsection{Experimental setting}
The hyperparameters used in the simulation are listed in Table \ref{table:1}. The reward for each step of an episode is considered as $r(\mathbf{e},\mu_{\theta}(\mathbf{e}))=-e_1^2 -e_2^2$. To penalize the reward at each step equally, $\gamma=1$ is considered in the simulation studies. This is a reasonable assumption since the controller design procedure is considered as an episodic task (for non-episodic tasks $\gamma<1$ should be chosen to avoid unlimited return). The goal of using the DDPG network is for the states of the error system to reach zero in the desired time horizon (here, the horizon is considered to be $T=7\,s$). By assuming $t_s=0.1\,s$, the prediction horizon (episode length) $N$ is $70$. Besides, if the reward at each time step exceeds $-20$, the corresponding episode during the learning phase will be terminated.
Each episode begins from the initial states $[0,0]^T$.

\begin{table}
\caption{Simulation hyperparameters}
\centering
\begin{tabular}{|p{1.3cm}|p{1.3cm}||p{1.3cm}|p{1.3cm}|} 
 \hline
 parameter & value & parameter & value\\ [0.5ex] 
 \hline\hline
 $\alpha_a$ & $10^{-4}$
 & $\alpha_c$ & $5\times 10^{-3}$ \\ 
 $\gamma$ & $1$ & $\tau$ & $5\times 10^{-3}$\\
 $N$ & $70$ & $|B_1|$ & $70$\\
 $\underline{G}$ & $-20$ & & \\[1ex]
 \hline
\end{tabular}
 \label{table:1}
\label{table1}
\end{table}


\subsubsection{Results and discussion} 
\begin{figure}[!htp]
\begin{center}
\includegraphics[width=1\linewidth]{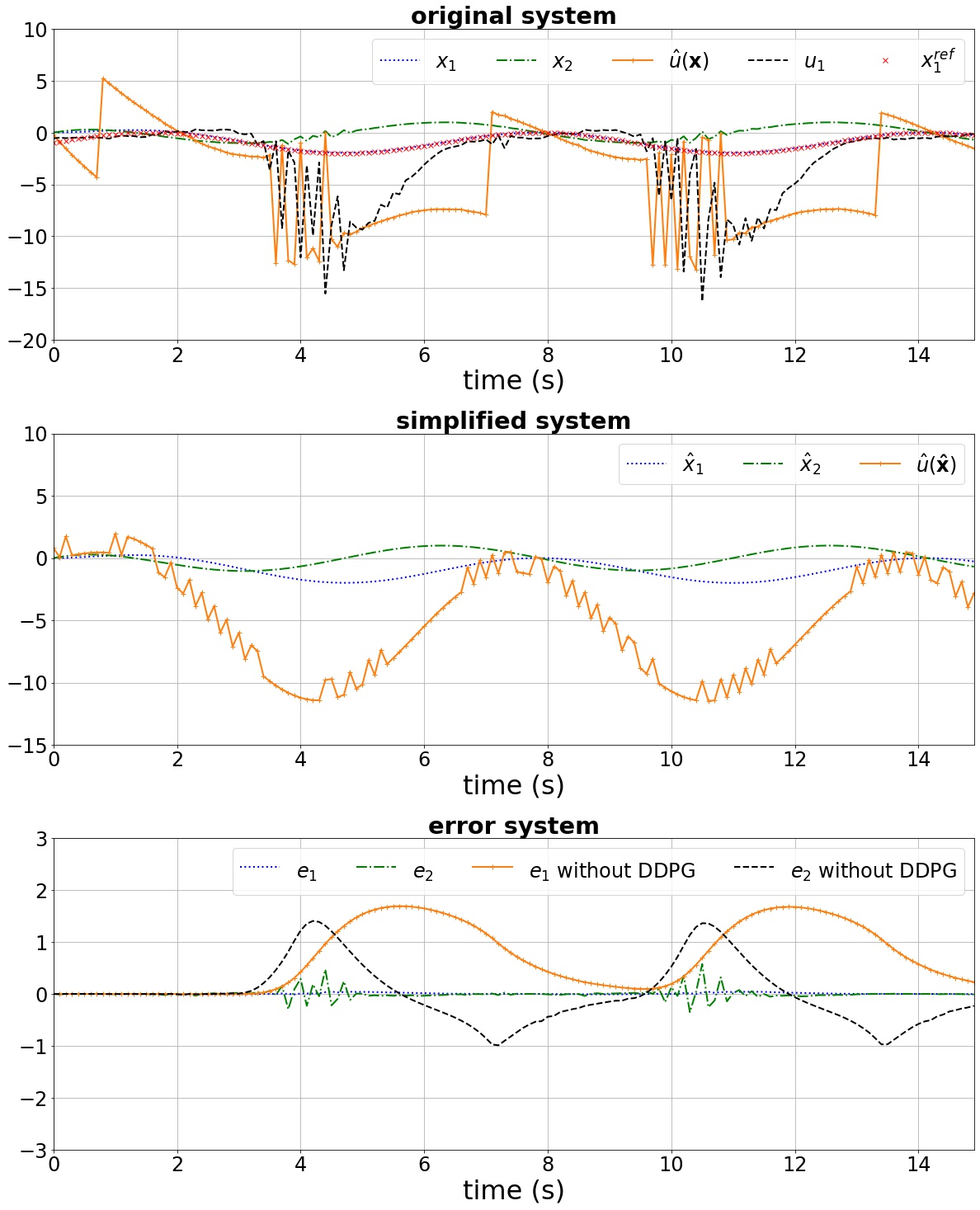}    
\caption{Closed-loop system performance, where the proposed SMC is used to control the original system in \eqref{example_original}. The first subplot shows original system states and system input ($u=\hat{u}_{eq}+u_1$). The second subplot shows simplified system states, while the last subplot depicts error system states. The goal of the proposed SMC is for the states of the error system to converge to zero quickly. } 
\label{fig:performance}
\end{center}
\vspace{-15pt}
\end{figure}

\begin{figure}[!htp]
\begin{center}
\includegraphics[width=1\linewidth]{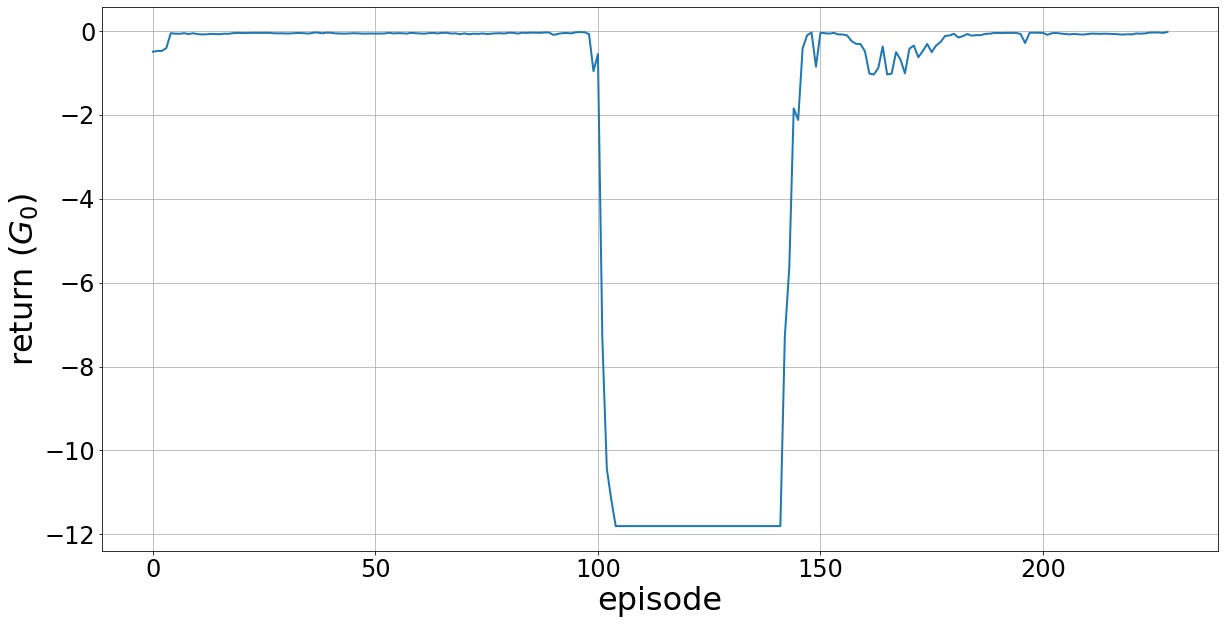}    
\caption{Return ($G_0$) vs. episodes: after about 175 episodes, DDPG learns the suitable action for the system.} 
\label{fig:reward}
\end{center}
\vspace{-15pt}
\end{figure}

The performance of the proposed controller after convergence is shown in Fig. \ref{fig:performance}. The first subplot shows the original system states, the tracking signal $x_1^{ref}$, the control law calculated using the available simplified model (model-based controller $\hat{u}(x_1,x_2)$), and the output of the DDPG network ($u_1$). The second subplot shows the performance of the simplified system using the SMC controller $\hat{u}(\hat{x}_1,\hat{x}_2)$. The last subplot depicts the error system dynamics for two cases: 1) network is employed to compensate for the unknown parts in the original system dynamics; 2) when only the control law calculated based on the simplified system is used. Simulation results depict the efficacy of the proposed controller design in stabilizing the error system dynamics. It is noted that, in this example, the goal is to track a specific reference ($x^{ref}_1$), and the controller is successfully able to track the reference (in other words, $e_1$ converges to zero). The results reveal the capability of the proposed method to control a partially-known system, in which not only the dynamics are not completely available but also the available knowledge is not accurate (here the constants $\hat{c},\,\hat{k},\,\hat{m}$ do not match their real values). Fig. \ref{fig:reward} shows the return ($G_0$) at each episode during the learning process; as observed, after about 175 episodes, the proper action is found.

\begin{figure}[!htp]
\begin{center}
\includegraphics[width=1\linewidth]{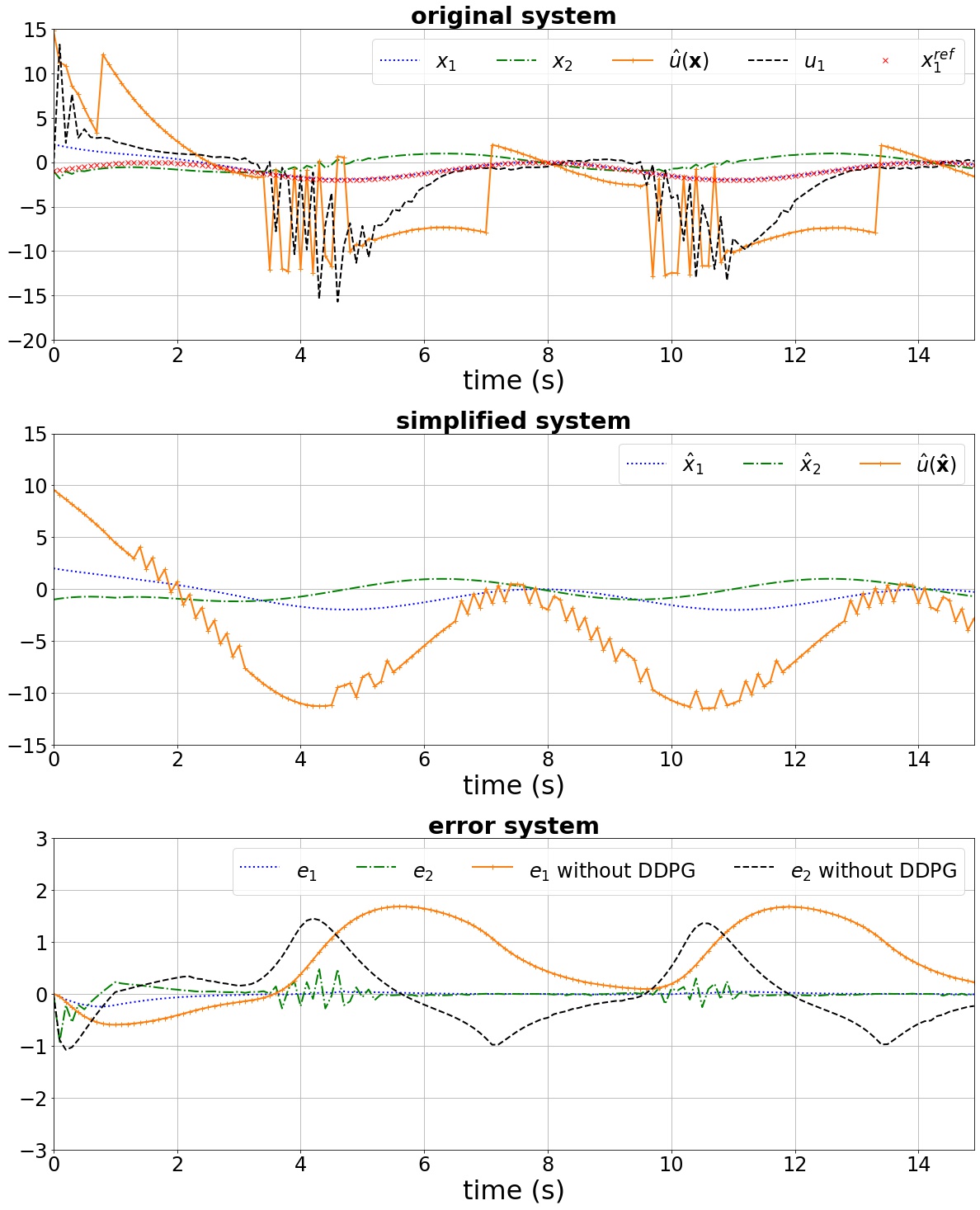}    
\caption{Evaluation of the generalization capability of the proposed controller. The controller learns a suitable control law beginning from the state $[0\,\,\,\,0]^T$. However, the controller shows a strong performance even with a non-zero initial condition; in this simulation, the initial condition is $[2\,\,-1]^T$.} 
\label{generalization}
\end{center}
\vspace{-6mm}
\end{figure}

To demonstrate the generalization capability of the proposed controller, we use the learned controller to evaluate its performance when the initial condition for the system changes. The system is trained with $[0\,\,\,\,0]^T$ as the initial state, while the performance is evaluated when the system initial condition is $[2\,\,-1]^T$. From the results shown in Fig. \ref{generalization}, it is observed that with the proposed control design approach, successful tracking of the reference is achieved although the initial state for evaluation is different from the one used for learning the controller.

\section{Conclusion}

\noindent In this paper, model-based and data-driven control design approaches were fused to build a sliding mode controller for a class of partially-known nonlinear systems. A deterministic policy gradient approach (known as DDPG) was employed to cope with the mismatch between the available model of the system and the actual system dynamics online. A procedure for designing such controller was proposed and the performance of the design approach was evaluated through simulation studies.

\bibliographystyle{ieeetr}
\bibliography{references}            

\end{document}